\documentclass[a4paper]{article}
\usepackage{ISCSLP2026}
\usepackage{iftex}
\ifPDFTeX\else
  \usepackage{fontspec}
\fi
\usepackage{ifthen}
\newboolean{blind}
\setboolean{blind}{false}

\usepackage{cite}
\usepackage{amsmath,amssymb,amsfonts}
\usepackage{algorithm}
\usepackage{algpseudocode}
\usepackage{graphicx}
\usepackage{booktabs}
\usepackage{multirow}
\usepackage{textcomp}
\usepackage{xcolor}
\usepackage{url}

\begin{document}

\title{Context-Aware ASR for Mandarin Technical Lectures}

\name{
	\ifthenelse{\boolean{blind}}{Anonymous to ISCSLP}
	{Ho-Lam Chung$^{1,2}$, Yiming Chen$^{2}$, Hung-yi Lee$^{1}$}
}

\address{
  \ifthenelse{\boolean{blind}}{Anonymous to ISCSLP}
  {
  	$^1$National Taiwan University, Taiwan \quad $^2$ASUS
  }
}

\email{
	\ifthenelse{\boolean{blind}}{Anonymous to ISCSLP}
	{holam.chung@protonmail.com, yiming.chen@u.nus.edu, tlkagkb93901106@gmail.com}
}

\maketitle

\begin{abstract}
Technical lectures mix Mandarin speech with English technical terms. These terms carry the core meaning of the lecture, yet they occupy few characters. Character error rate (CER) therefore hides their recognition failures. We study whether lecture context helps recognize these terms. We build a term-rich Mandarin AI/ML lecture benchmark, and we define term-centric metrics that measure technical-term recognition directly. We then propose a two-pass, reference-free decoding method. The first pass runs segment-only ASR. We extract the most frequent technical terms from the first-pass hypotheses, and we prompt the recognizer with this self-built glossary in the second pass. Across five ASR backbones, the first-pass glossary raises term recall for every model and holds or lowers CER on all five. On Breeze-ASR-25 it lifts term recall from 52.50\% to 60.13\% while lowering CER, and a hybrid that adds a small external term list reaches 62.05\% recall and 82.73\% term precision. Lecture context, recovered from the model's own output, is a practical signal for technical-term recognition. Term-centric evaluation exposes errors that CER misses.
\end{abstract}

\section{Introduction}

Lecture ASR powers subtitles, search, and note-taking for students. In AI/ML lectures, bilingual technical terms anchor the content. Terms such as \textit{RAG}, \textit{SWE-bench}, \textit{AI Agent}, \textit{token}, and \textit{embedding} name the concepts the speaker explains. A student who reads ``RIG'' instead of ``RAG'' loses the point of the segment.

These terms occupy a small fraction of the characters. A transcript can reach low CER while still misrecognizing the key term. The Mandarin around the term is fluent, so CER stays low. The usable content, however, is wrong. CER does not capture this failure.

Lecture terms are bursty. Once the speaker introduces \textit{AI Agent}, the term recurs across many later segments. The lecture title, the previous transcript, and the repeated terms all point to the same vocabulary. Segment-only decoding discards this signal. It recognizes each segment in isolation, so it cannot use the strong context that the lecture provides.

We address this gap by building a term-rich benchmark, measuring term recognition directly, and supplying a self-built glossary as decoding context.

Our contributions are as follows.
\begin{itemize}
\item We release a term-rich Mandarin AI/ML lecture ASR benchmark. It contains 8{,}888 technical-term occurrences over a 5.01-hour code-switched test set.
\item We define term-centric evaluation. It reports term recall, term precision, term F1, and term error rate, so it separates technical-term accuracy from overall CER.
\item We propose a two-pass, reference-free glossary prompt. It extracts a lecture glossary from first-pass hypotheses, and it improves term recall across five ASR backbones while holding or lowering CER on all of them. A hybrid that adds a small external glossary reaches 62.05\% term recall on Breeze-ASR-25, closing part of the remaining headroom.
\end{itemize}

\section{Related Work}

\textbf{Code-switching Mandarin-English ASR.} Mandarin-English code-switching is a long-standing challenge for ASR. Public corpora such as SEAME~\cite{lyu2010seame}, the ASRU 2019 challenge set~\cite{shi2020asru}, and ASCEND~\cite{lovenia2022ascend} established Mandarin-English benchmarks, but they target conversational rather than lecture speech. Recent work fine-tunes large models on Mandarin with English terms~\cite{chou2025selfrefining}, and distills language-specific knowledge for code-switching~\cite{tseng2024codeswitch}. These methods improve the acoustic model. We instead keep the model frozen, and we add context at decoding time.

\textbf{Contextual biasing.} Contextual ASR injects external knowledge into recognition. Deep context learns to bias toward a phrase list~\cite{pundak2018deepcontext}. Trie-based deep biasing and shallow fusion bias streaming recognizers toward a catalog~\cite{le2021contextual}. Audio-language models couple speech encoders with LLMs~\cite{chu2023qwenaudio,tang2024salmonn}, and LLM-based recognizers consume free-form context~\cite{bai2024seedasr}. Qwen3-ASR treats context tokens as background knowledge for biasing~\cite{qwen2026asr}. These methods assume a context source is available. We build the context from the model's own first-pass output, so the method needs no external list.

\textbf{Prompting and retrieval.} Whisper accepts an initial prompt that conditions decoding~\cite{radford2023whisper}. Two-pass ASR refines first-pass output, either by rescoring streamed hypotheses with a second decoder~\cite{sainath2019twopass,hu2020deliberation} or by correcting the first-pass N-best list with an LLM~\cite{chen2023hyporadise}. Retrieval-augmented generation supplies retrieved evidence to a generator~\cite{lewis2020rag}. Our two-pass glossary is an in-domain analogue: the first pass retrieves over the lecture itself, and the second pass conditions on the retrieved terms. Unlike rescoring or correction, we feed the terms back as context rather than editing the hypotheses, and we derive that list from the model's own lecture-level output, so the method is reference-free.

\section{Benchmark}

We build the benchmark from a publicly available Mandarin-English AI/ML lecture series\footnote{Dataset, code, and term metadata will be released upon publication.}. The lectures are spontaneous speech, with English technical terms code-switched into Mandarin. We segment the lectures and pair each segment with its reference transcript.

We then build a term-rich test set. A rule-based recognizer extracts technical terms from the reference transcripts. It targets four kinds of terms: acronyms such as \textit{AI} and \textit{LLM}; model and tool names such as \textit{Llama}, \textit{Whisper}, and \textit{OpenClaw}; code-like strings such as \textit{SWE-bench} and \textit{log1.txt}; and multi-word terms such as \textit{AI Agent} and \textit{Context Engineering}. We canonicalize each term with NFKC normalization, lowercasing, and light stemming, so surface variants map to one key. A segment enters the term-rich set when it contains at least one term.

\begin{figure}[t]
\centering
\includegraphics[width=\columnwidth]{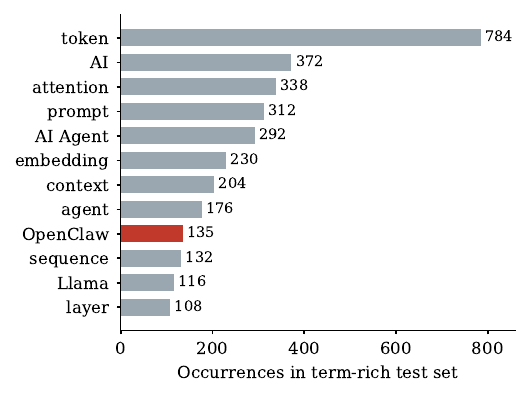}
\caption{Technical-term frequency in the term-rich test set, top 12 terms. The distribution is heavy-tailed. The lecture-specific tool name \textit{OpenClaw} (red) occurs 135 times, which shows that even a name the model has never seen repeats heavily within one lecture.}
\label{fig:burst}
\end{figure}

Table~\ref{tab:bench} summarizes the benchmark. The full set has 29{,}043 segments and 16.92 hours. The term-rich test set has 7{,}443 segments and 5.01 hours, with 8{,}888 term occurrences and 1{,}030 unique terms over 15 lectures. Terms are bursty. Fig.~\ref{fig:burst} shows the term frequencies. The term \textit{token} occurs 784 times, \textit{AI Agent} 292 times, and the lecture-specific name \textit{OpenClaw} 135 times. This burstiness is the signal that motivates lecture-level context. The distribution is heavy-tailed: acronyms and short model names lead the head, while multi-word terms and code-like strings fill the tail. A method must handle both the frequent acronyms and the rare composite names.

\begin{table}[t]
\centering
\caption{Benchmark statistics. The term-rich test set isolates segments that contain at least one technical term.}
\label{tab:bench}
\small
\begin{tabular}{lr}
\toprule
\textbf{Item} & \textbf{Value} \\
\midrule
Full set segments & 29{,}043 \\
Full set duration & 16.92 h \\
Term-rich segments & 7{,}443 \\
Term-rich duration & 5.01 h \\
Term occurrences & 8{,}888 \\
Unique canonical terms & 1{,}030 \\
Lectures & 15 \\
\bottomrule
\end{tabular}
\end{table}

\section{Term-Centric Evaluation}

CER measures character accuracy over the whole segment. It does not tell us whether the recognizer captured the technical terms. We therefore add four term-centric metrics. We compute them over canonical term keys.

We define \textit{term recall} as the fraction of reference term occurrences whose canonical key appears in the hypothesis of the same segment. We define \textit{term precision} as the fraction of predicted term keys that appear in the reference. We define \textit{term F1} as their harmonic mean. We define \textit{term error rate} as the edit distance between the reference and hypothesis term sequences, normalized by the reference term count.

We keep CER as the overall accuracy measure. We normalize all Mandarin text with OpenCC\footnote{\url{https://github.com/BYVoid/OpenCC}} (\texttt{s2twp}) and apply NFKC and lowercasing before scoring. A usable transcript needs both high recall and high precision, and CER captures neither. The two metrics can move in opposite directions: a recognizer that copies a long context into its output raises recall while collapsing precision, as Table~\ref{tab:context} shows. We therefore report recall and precision together, and treat term recall as the headline metric because it counts how many technical terms the recognizer recovered.

Table~\ref{tab:main} reports five recognizers under segment-only decoding. Breeze-ASR-25 has the best CER and the best term error rate. Yet it recovers only 52.50\% of the terms. Whisper-large-v3-turbo has slightly higher term recall but much lower term precision. Every model leaves a large share of technical terms unrecognized. A low CER does not imply usable technical transcripts.

\begin{table}[t]
\centering
\caption{Segment-only baselines. CER is computed on the full 16.92\,h set; term metrics on the 5.01\,h term-rich subset. All values are percentages. \textbf{Bold} marks the best column value, \underline{underline} the second best. Lower is better for CER and TermER; higher is better otherwise.}
\label{tab:main}
\resizebox{\columnwidth}{!}{%
\begin{tabular}{lrrrrr}
\toprule
\textbf{Model} & \textbf{CER} & \textbf{Recall} & \textbf{Prec.} & \textbf{F1} & \textbf{TermER} \\
\midrule
Breeze-ASR-25 & \textbf{10.72} & \underline{52.50} & \textbf{80.55} & \textbf{63.57} & \textbf{48.66} \\
whisper-l-v3-turbo & \underline{11.95} & \textbf{53.90} & 65.11 & \underline{58.98} & \underline{49.67} \\
Qwen3-ASR-1.7B & 15.28 & 49.03 & \underline{72.84} & 58.61 & 53.31 \\
Qwen3-ASR-0.6B & 16.64 & 44.28 & 71.19 & 54.60 & 58.20 \\
Breeze-ASR-26 & 24.63 & 27.31 & 56.26 & 36.77 & 75.14 \\
\bottomrule
\end{tabular}
}
\end{table}

\begin{table*}[t]
\centering
\renewcommand{\arraystretch}{0.95}
\caption{Contextual decoding across five ASR backbones on the 5.01\,h term-rich set. CER is rescored on this subset, so baselines match the contextual rows. $\Delta$R is the term-recall gain over the baseline. $\star$ marks the proposed reference-free prompt. \textbf{Bold} marks the best value per column within each model. Values are percentages; lower is better for CER and TermER.}
\label{tab:context}
\resizebox{\textwidth}{!}{%
\begin{tabular}{llrrrrrr}
\toprule
\textbf{Model} & \textbf{Context} & \textbf{CER} & \textbf{Recall} & \textbf{$\Delta$R} & \textbf{Prec.} & \textbf{F1} & \textbf{TermER} \\
\midrule
\multirow{3}{*}{Breeze-ASR-25}
 & baseline (segment-only) & 11.37 & 52.50 & -- & 80.55 & 63.57 & 48.66 \\
 & + first-pass glossary ($k${=}30) $\star$ & 9.79 & \textbf{60.13} & +7.63 & 81.20 & \textbf{69.09} & \textbf{41.38} \\
 & + title + prev-ASR + glossary & \textbf{9.19} & 59.51 & +7.01 & \textbf{82.19} & 69.03 & 41.74 \\
\midrule
\multirow{3}{*}{whisper-l-v3-turbo}
 & baseline (segment-only) & 13.80 & 53.90 & -- & \textbf{65.11} & 58.98 & 49.67 \\
 & + first-pass glossary ($k${=}30) $\star$ & \textbf{12.82} & \textbf{60.86} & +6.95 & 63.79 & \textbf{62.29} & \textbf{45.93} \\
 & + title + prev-ASR + glossary & 15.89 & 59.74 & +5.84 & 65.03 & 62.28 & 48.77 \\
\midrule
\multirow{3}{*}{Qwen3-ASR-1.7B}
 & baseline (segment-only) & 16.13 & 49.03 & -- & 72.84 & 58.61 & 53.31 \\
 & + first-pass glossary ($k${=}30) $\star$ & 14.02 & 56.56 & +7.53 & 74.43 & 64.27 & 47.10 \\
 & + title + prev-ASR + glossary & \textbf{13.93} & \textbf{56.60} & +7.57 & \textbf{76.11} & \textbf{64.92} & \textbf{46.26} \\
\midrule
\multirow{3}{*}{Qwen3-ASR-0.6B}
 & baseline (segment-only) & 18.91 & 44.28 & -- & \textbf{71.19} & 54.60 & \textbf{58.20} \\
 & + first-pass glossary ($k${=}30) $\star$ & \textbf{18.86} & 53.74 & +9.45 & 65.92 & \textbf{59.21} & 54.64 \\
 & + title + prev-ASR + glossary & 25.39 & \textbf{54.43} & +10.15 & 59.79 & 56.99 & 64.99 \\
\midrule
\multirow{3}{*}{Breeze-ASR-26}
 & baseline (segment-only) & 38.28 & 27.31 & -- & 56.26 & 36.77 & 75.14 \\
 & + first-pass glossary ($k${=}30) $\star$ & 26.05 & 44.89 & +17.59 & 57.15 & 50.28 & 62.27 \\
 & + title + prev-ASR + glossary & \textbf{21.67} & \textbf{49.73} & +22.42 & \textbf{63.32} & \textbf{55.71} & \textbf{54.53} \\
\bottomrule
\end{tabular}
}
\end{table*}

\section{Method}

We propose a two-pass, reference-free decoding method. We call it \textsc{asr-glossary}. The idea is simple. A lecture repeats its terms, so the model's own first-pass output already contains those terms. We collect them, and we feed them back as context.

Algorithm~\ref{alg:method} states the procedure. The first pass runs segment-only ASR over the lecture. We extract canonical terms from every first-pass hypothesis, and we rank them by frequency across the lecture. We keep the top $k$ surface forms as the first-pass glossary. The second pass re-decodes each segment with this glossary as context.

\begin{algorithm}[t]
\caption{Two-pass glossary-prompted decoding}
\label{alg:method}
\begin{algorithmic}[1]
\Require lecture segments $S=\{s_1,\dots,s_n\}$, recognizer $M$, glossary size $k$
\State \textbf{Pass 1:} $h_i \gets M(s_i)$ for all $s_i$ \Comment{segment-only}
\State $C \gets \textsc{Counter}()$
\For{each $h_i$}
  \For{each term $t \in \textsc{ExtractTerms}(h_i)$}
    \State $C[t] \mathrel{+}= 1$
  \EndFor
\EndFor
\State $G \gets$ top-$k$ surface forms of $C$ by frequency
\State \textbf{Pass 2:} $h'_i \gets M(s_i \mid G)$ for all $s_i$ \Comment{glossary as context}
\Ensure refined transcripts $\{h'_1,\dots,h'_n\}$
\end{algorithmic}
\end{algorithm}

We set $k=30$. A short glossary keeps the prompt compact, so it avoids over-conditioning. Frequency ranking captures the bursty repeats, so the glossary holds the terms that matter for this lecture. A sweep over $k$ (Sec.~\ref{sec:analysis}) confirms that $k=30$ gives the lowest CER, and a selection ablation confirms that lecture-level frequency, not the mere presence of a term, is the signal that matters. We use the surface form, so the prompt reads like natural text.

\textbf{Conditioning interface.} The two model families consume the glossary differently. For Whisper-family models, including Breeze, we pass the glossary as an initial prompt through prompt tokens. We apply it to the first decoding window, we cap it at 160 tokens, and we decode greedily. For Qwen3-ASR, we pass the glossary through the native context field as a system message. The instruction tells the model to use the context for technical-term disambiguation and not to copy unspoken text. We call this last constraint the guard sentence. Qwen3-ASR is trained to treat such context as background knowledge~\cite{qwen2026asr}, so the design matches the model.

\textbf{Context variants.} We compare the first-pass glossary against five alternative context sources and an oracle upper bound.
\begin{itemize}
\item \textbf{Title}: the lecture title only.
\item \textbf{Prev-ASR}: the previous five first-pass segments.
\item \textbf{Combined}: joins the lecture title, prev-ASR segments, and first-pass glossary.
\item \textbf{External list}: a fixed, course-level ML/LLM term list, ranked by lecture-title match.
\item \textbf{Hybrid}: fills the first 15 prompt slots from the external list and the remaining slots from the first-pass glossary.
\item \textbf{Oracle}: prompts with terms from the reference transcript. Not deployable; reported as an upper bound only.
\end{itemize}
All variants except the oracle are reference-free.

\section{Experiments}

\textbf{Setup.} We decode five backbones with context, and we run the glossary ablations on Breeze-ASR-25. We rescore each segment-only baseline on the 5.01-hour term-rich subset, so its CER is comparable to the contextual rows. These subset CERs are higher than the full-set CERs in Table~\ref{tab:main}, because the term-rich subset is harder. We decode greedily unless stated otherwise. We test significance with a paired bootstrap over segments, using 2{,}000 resamples. All term-recall gains below are significant at $p<0.001$~\cite{koehn2004significance}.

\textbf{Main result.} Table~\ref{tab:context} reports the contextual results. The first-pass glossary raises term recall for all five backbones. The gain ranges from +6.95 on whisper-large-v3-turbo to +9.45 on Qwen3-ASR-0.6B across the four competitive backbones, and it reaches +17.59 on the weak Breeze-ASR-26. The glossary also holds or lowers CER on all five models. On Breeze-ASR-25 it cuts CER from 11.37\% to 9.79\% while lifting recall from 52.50\% to 60.13\%.

\begin{table}[t]
\centering
\caption{Term-source comparison on Breeze-ASR-25 over the 5.01\,h term-rich set. \textit{First-pass glossary}: terms extracted from the model's own first-pass output. \textit{External list}: a fixed course-level ML/LLM term list. \textit{Hybrid}: first 15 slots from the external list, rest from the first-pass glossary. \textit{Oracle}: terms from the reference transcript (upper bound, not deployable). $\star$ marks reference-free settings. \textbf{Bold} marks the best deployable value per column. All values are percentages; lower is better for CER and TermER.}
\label{tab:source}
\resizebox{\columnwidth}{!}{%
\begin{tabular}{lrrrrr}
\toprule
\textbf{Source} & \textbf{CER} & \textbf{Recall} & \textbf{Prec.} & \textbf{F1} & \textbf{TermER} \\
\midrule
baseline (segment-only) & 11.37 & 52.50 & 80.55 & 63.57 & 48.66 \\
first-pass glossary $\star$ & 9.79 & 60.13 & 81.20 & 69.09 & 41.38 \\
external list $\star$ & 10.04 & 60.29 & 81.89 & 69.45 & 41.38 \\
hybrid $\star$ & \textbf{9.40} & \textbf{62.05} & \textbf{82.73} & \textbf{70.91} & \textbf{39.39} \\
\midrule
\textit{oracle (reference)} & \textit{8.19} & \textit{68.88} & \textit{86.70} & \textit{76.77} & \textit{32.71} \\
\bottomrule
\end{tabular}
}
\end{table}

\textbf{Combined context vs.\ first-pass glossary.} The combined context (title + prev-ASR + glossary) helps some models and hurts others. It gives the best recall on Qwen3-ASR-1.7B and on the weak Breeze-ASR-26, but it raises CER on whisper-large-v3-turbo to 15.89\% and on Qwen3-ASR-0.6B to 25.39\%. The added length is the cause. A small model cannot manage long context, so the extra text corrupts decoding. The raw prev-ASR context is the worst case: on Qwen3-ASR-0.6B it alone drives CER to 34.01\%, because the model copies whole mis-recognized sentences. The first-pass glossary avoids this failure. It is short, and it improves recall on every model. The glossary is therefore the robust first choice, and the combined context is a per-model option.

\textbf{Closing the headroom.} The oracle sits well above every deployable setting, at 68.88\% recall on Breeze-ASR-25, 64.41\% on Qwen3-ASR-1.7B, and 62.84\% on Qwen3-ASR-0.6B. The gap is mostly a coverage problem. Of the terms only the oracle recovers, 55\% to 65\% never appear in any first-pass hypothesis. These are lecture-specific names: \textit{OpenClaw} is heard as ``open cloud'', \textit{ACON} as ``Aiken'', and \textit{AppWorld} as ``app world''. A first-pass glossary cannot recover them, so we add a small external list of 15 course-level terms per lecture, which still uses no reference transcript. Table~\ref{tab:source} compares the sources on Breeze-ASR-25. The external list alone matches the first-pass glossary. The hybrid of the two reaches 62.05\% recall and 9.40\% CER, a significant +1.92 recall gain over the first-pass glossary. Relative to the segment-only baseline, the hybrid improves all four term metrics at once: it raises recall from 52.50\% to 62.05\% and precision from 80.55\% to 82.73\%, lifts F1 from 63.57\% to 70.91\%, and lowers CER from 11.37\% to 9.40\%, each significant at $p<0.001$. The gain is therefore not a recall-for-precision trade. It is the strongest deployable setting we test.

\subsection{Analysis}
\label{sec:analysis}

\textbf{Glossary size.} We sweep the glossary size on Breeze-ASR-25 over $k \in \{10,20,30,50,100,\text{all}\}$. Recall rises with $k$ and saturates near $k{=}100$, since each lecture yields at most 106 first-pass candidates. CER is lowest at $k{=}30$ (9.79\%), and term F1 peaks at $k{=}50$ (69.50\%). Beyond $k{=}50$, the recall gain shrinks while CER rises and precision drops. We keep $k{=}30$ as a compact operating point that gives the lowest CER and near-peak F1.

\textbf{Term-selection criterion.} We fix $k{=}30$ and change how the 30 terms are chosen. Frequency ranking gives the best recall (60.13\%), F1, and CER. First-occurrence (57.27\%) and random (58.58\%) selection are significantly worse. Downweighting generic terms stays close on recall (59.90\%) but still lowers F1 and precision. The bursty repeat count, not the mere presence of a term, is what makes the glossary work.

\textbf{Where it helps.} The recall gain spans all four term kinds (Table~\ref{tab:category}). It is concentrated on acronyms and model names for Breeze-ASR-25, and on acronyms and multi-word terms for the small Qwen3-ASR-0.6B, so a weak baseline has the most term headroom to recover. Every per-category gain is significant under paired bootstrap. For Qwen3, the guard sentence protects precision: removing it lowers precision on Qwen3-ASR-0.6B by 1.39 points and raises its CER.

\begin{table}[t]
  \caption{Per-category term-recall gain (percentage points) of the first-pass glossary over segment-only decoding, on the 5.01 h term-rich set, split by the four reference term kinds. All gains are significant under paired bootstrap ($p<0.01$).}
  \label{tab:category}
  \centering
  \resizebox{\columnwidth}{!}{%
  \begin{tabular}{l cccc}
    \toprule
    \textbf{Model} & \textbf{Acronym} & \textbf{Model/Name} & \textbf{Code} & \textbf{Multi-word} \\
    \midrule
    Breeze-ASR-25   & $+13.37$ & $+9.31$ & $+6.03$ & $+4.08$ \\
    Qwen3-ASR-1.7B  & $+10.54$ & $+5.70$ & $+6.60$ & $+10.04$ \\
    Qwen3-ASR-0.6B  & $+15.68$ & $+5.50$ & $+1.97$ & $+14.89$ \\
    \bottomrule
  \end{tabular}%
  }
\end{table}

\textbf{Robustness checks.} Under beam search ($b{=}5$) the recall gain holds at +7.56 and CER is non-worse, so it is not a greedy-decoding effect. The first-pass glossaries are clean: weighted purity is 95.56\% on Breeze-ASR-25, 92.19\% on Qwen3-ASR-1.7B, and 90.75\% on Qwen3-ASR-0.6B, so the reference-free list is mostly real lecture terms. A third decoding pass that rebuilds the glossary from the second-pass output raises recall by a further +0.90 to 61.03\%, but it does not improve F1, CER, or term error rate, because precision falls as the glossary grows. One refinement pass is therefore enough. The glossary also rescues near-miss forms: ``aia'' becomes \textit{AI agent}, ``lnn'' becomes \textit{llm}, ``pump'' becomes \textit{prompt}, and ``SW1 bench'' becomes \textit{SWE-bench}.

\section{Conclusion}

We studied ASR for Mandarin technical lectures, where CER hides technical-term failures. We measured term recognition directly with term recall, precision, F1, and term error rate, and we proposed a two-pass, reference-free glossary prompt built from first-pass hypotheses. It raises term recall on all five backbones while holding or lowering CER, and the lecture-level frequency of the terms, not their mere presence, drives the gain. On Breeze-ASR-25 a hybrid with a small external list reaches 62.05\% term recall and 9.40\% CER.

The method reframes contextual biasing as retrieval over the recording itself: the first pass retrieves, and the second pass conditions on the retrieved terms with no external catalog. Because term burstiness drives the gain, it should carry to other long-form code-switched audio such as meetings and seminars. The benchmark and the term-centric metrics also transfer: any domain with specialized vocabulary and low-CER models would benefit from evaluation that separates term accuracy from overall character accuracy.

Several limitations remain. The benchmark covers one instructor and one domain (AI/ML). Multi-speaker and cross-domain evaluation, for example on medical or legal lectures, would test whether the burstiness assumption generalizes. The term extractor is rule-based, so it misses terms that do not match pre-defined patterns. A learned extractor could improve both benchmark coverage and first-pass glossary quality. The largest open gap is coverage: 55\% to 65\% of oracle-only terms never appear in any first-pass hypothesis, because the model has never seen the surface form. Closing this gap may require retrieval from course-level materials such as slides and syllabi, which are typically available in educational settings.

\bibliographystyle{IEEEtran}
\bibliography{refs}

\end{document}